\documentstyle[preprint,revtex]{aps} 
\def\al#1{\mbox{$\alpha_ #1 ^{-1}$}}

\def\g#1{\mbox{$g_{#1}^2$}}
\def\e{\mbox{$e^2$}}
\def\tg{\mbox{$\tilde{g_2}^2$}}
\def\te{\mbox{$\tilde{e}^2$}}
\def\l#1{\mbox{$\ln \frac{#1}{\mu^2}$}}
\def\ep{\mbox{$\displaystyle{\frac{2}{\varepsilon}}$}}

\begin{document}

\draft

\preprint
{
\begin{flushright}
LAEFF--93/012\\
September 1993
\end{flushright}
}

\begin{title}
Threshold Effects and Perturbative Unification \\
\end{title} 
\author{M. Bastero--Gil$^{a,}$\cite{aa1}, V. Man\'{\i}as$^b$, 
J. P\'erez--Mercader$^{a,}$\cite{aa2}}
\begin{instit}  
$^{(a)}$Laboratorio de Astrof\'{\i}sica Espacial y F\'{\i}sica
Fundamental\\
Apartado 50727, 28080 Madrid, Spain
\end{instit}
\begin{instit}
$^{(b)}$Departamento de F\'{\i}sica, Facultad de C. Exactas, UNLP\\
C.C. 67, 1900 La Plata, Argentina
\end{instit}
\begin{abstract} 
 We discuss the effect of the renormalization procedure in the computation    
of the unification point for  running coupling constants. We explore the
effects  of  threshold--crossing on the $\beta$--functions.  We compute the
running of  the  coupling constants of the Standard Model, between $m_Z$ and
$M_P$,  using  a mass dependent subtraction procedure, and then compare the
results with $\overline{MS}$, and with the $\theta$-- function approximation. 
We also do this for the Minimal Supersymmetric extension of the Standard 
Model. In the latter,  the bounds on susy masses that one obtains by
requiring  perturbative unification are dependent, to some extent, on the
procedure. 
\end{abstract} 
\narrowtext
\vspace{.2in}

\pagebreak

The last years have seen a revival of the study of scenarios for a  possible 
perturbative unification of the 
strong, electromagnetic and weak interactions into a simple group $G$, the
simplest candidate being $SU(5)$ \cite{glashow,quinn}.
 In this way, the three couplings of the Standard Model 
(SM), $\,g_3,g_2, g'$, for $SU(3)_c\times SU(2)_L\times U(1)_Y$, would evolve 
up to a common value $g$ at  the unification scale  $M_X$ (order
of $10^{16}\, GeV$) from their disparate value at $m_Z$. Thus the study of the 
evolution  of these couplings from low
to high energies, can be used to obtain information about the possible 
validity of the perturbative unification  picture.

Since we pretend to compare experimental values with theoretical
predictions for the coupling constants $g_i$, how to actually study the scale 
evolution of the couplings
becomes an issue.  The tool to carry this out is the renormalization group 
(RG), with  $\beta_i$--functions  defined as 
$\beta_i=\frac{d\,g_i}{d\ln\mu}$,  from which we can get
$g_i(\mu)$, $\mu$ being the energy scale at which we probe the system, and
$g_i$ the  renormalized coupling. The physical coupling constants are
measured  at the laboratory at a  given  energy scale  $\mu_0$, 
typically of order of $m_Z$. We {\it identify} these values  with
$g_i(\mu_0)$, and with the help of the RGE we infer the evolution   of the 
couplings as the scale  changes.
 However,  in order to compute the $\beta_i$--functions, one needs to specify a
renormalization {\it scheme}, and beta functions differ for one scheme to
another. Broadly speaking, one has  generally two choices of scheme at his/her
disposal:
 the modified minimal  subtraction scheme ($\overline{MS}$) \cite{ms},
and mass dependent subtraction procedures (MDSP) \cite{georgi}. In 
$\overline{MS}$ schemes,  the $\beta_i$--functions depend on the particle 
content at the energy scale on which one computes them and not explicitly on 
the masses nor  energy scale. On the other hand, MDSP schemes take into 
account a dependence on particle masses: each graph including particles with 
mass  $m_i$ running in the  loop, contains a function of the  ratio
$\mu^2/m_i^2$ associated  with it, where $\mu$ is the energy scale. The
decoupling  theorem \cite{appel} ensures that the contributions from
graphs with heavy field loops  are suppressed at low momenta \linebreak 
(much smaller than the masses involved).
 Naturally, the  functions $f(\mu^2/m_i^2)$  associated with each graph have 
the property that  $f(\mu^2/m_i^2) \rightarrow 0$, when $\mu^2/m_i^2
\rightarrow 0$, and  therefore the decoupling of the heavy  degrees of
freedom is explicitly  enforced. Because of their origin,  these functions
include the threshold effects arising as  $\mu$ becomes larger than the
putative particle mass.

 In addition to the above schemes, one can approximate the threshold
effects by Heaviside--$\theta$ functions and impose this on  the
$\overline{MS}$ $\beta$--functions, which are then integrated to give the 
effective couplings. Hence  a massive particle only contributes at scales 
larger than its  mass \cite{marciano}. However, this is only  an 
$approximation$ to the contribution given by the functions  $f(\mu^2/m_i^2)$ 
in the MDSP, and the results for the effective gauge couplings can be 
appreciably different (the more thresholds are crossed, 
the more different are the effective couplings).

 To $illustrate$ the relevance of the scheme--choice we calculate
$\alpha_{e}(m_W)$, integrating the  $\beta$--function from $m_e$  with
$\alpha_e^{-1}(m_e)=137.036$, to $m_W=81\,GeV$  and  assuming $m_t\ge
m_Z$, using the  three methods, {\it viz.}, (a) $\overline{MS}$  without any 
reference to the thresholds (which of course gives extremely inaccurate results
 as it is shown below), (b) $\theta$--functions, and (c) MDSP. It is
straightforward  to see that:
$$
\alpha_e^{-1} \mid_{\overline{MS}} = 120.06 \;,\;\;
\alpha_e^{-1} \mid_{\theta} = 128.54 \;,\;\;
\alpha_e^{-1} \mid_{MDSP} =129.33 \;.
$$

 Because thresholds modify the derivatives of the couplings 
($\beta$--functions) and we
are integrating over many decades in momenta, the effect due to the decoupling 
of massive degrees of freedom 
$and$ the subtraction procedure could  be  important, and in any case
one cannot disregard the effects of the thresholds.

 A further uncertainty is introduced when making considerations relative to 
unification, since the choice of the unifying  group,  $G$, impacts on the
renormalization of the couplings. Let us assume that this group breaks into 
the SM at scale  $M_X$. Since  the particle content  of $G$ is different from  
the SM we will have, in  general, light fields (particles of the SM)  
and heavy fields (with masses  of the order of the  unification scale $M_X$). 
At low energies these  heavy fields  decouple from the theory. To take into
account this  decoupling in a proper way one can, $e.g.$, integrate out the
heavy fields from the action  \cite{das,hall}.  Carrying this out, the
simple unification condition  $g_i(M_X)= g(M_X)$, is modified into  
$g_i(\mu)^{-2} = g(\mu)^{-2}-\lambda_i(\mu)$,  
where  the functions $\lambda_i$ depend on the masses of the heavy fields,
and  the interpolating scale $\mu$ satisfies $m_i \ll \mu \ll M_X$. 
Again the $values$ of the masses are important.

We are interested in studying how the presence of different  
masses affects the running of the couplings
(including  threshold effects), as well as its consequences on a possible
unification scenario for the SM. Therefore we  will study  here the SM with a
mass dependent renormalization procedure (MDSP).  The  relevant masses for
the problem are $m_Z,{}m_W,{}m_t$ and $m_h$, the Higgs  mass. We do not
consider the remaining  fermion masses because they are much smaller than
$m_Z$, so that for scales $\mu\geq m_Z$ we can regard all the fermions as
being  massless except for the top quark.

 In the symmetric phase, when $\mu \gg m_Z$, the couplings are 
\g{3}, \g{2}, $g'^2$ of $SU_c(3)\times SU_L(2)\times U_Y(1)$;  when the 
symmetry  breaks down, the relevant couplings are instead  \g{3}, \g{2} and
\e, with the coupling  $g'^2$  is given as a function of \g{2} and \e{} by 
\begin{equation}
\frac{1}{g'^2}= \frac{1}{\e}- \frac{1}{\g{2}}\;\;.
\label{echarge}
\end{equation}
In the laboratory we can measure  \g{3},  \g{2}  and \e{}   but not $g'^2$. 
Therefore we will {\it calculate} the  evolution of these three couplings and
$infer$ the evolution of $g'^2$  from (\ref{echarge}).

 We need to compute the $\beta_i$--functions for \g{3}, \g{2} and \e {}
in an MDSP. In  general, a renormalized coupling constant, \g{r}, is
related to  the bare constant \g{0} by  $\g{r}=Z \g{0}$,  
where $Z$ is a product of renormalization constants. A choice of subtraction  
procedure as well as 
an arbitrary subtraction  point $\mu$, are implicit in the computation of
the $Z$'s. In an MDSP there are two contributions to  $Z$: the pole part,
typical of the $\overline{MS}$ procedure, {\it and} finite contributions which 
depend on
both the masses of the virtual degrees of freedom and the  subtraction
point. Since  the $\beta$--function is defined as 
\begin{equation}
\beta=\g{r}\frac{d\ln\,\g{r}}{d\ln\,\mu}=\g{r} \frac{d\ln\,Z}{d\ln\,\mu}\;\;,
\end{equation}
this dependence on the masses and the subtraction point carries over to
the $\beta$--functions. Now, for a general gauge theory which includes
fermions and scalars, extra complications arise: 
there will be different  kinds of vertices  involving the coupling constants 
(fermion--boson and scalar--boson vertex in the case of abelian gauge
theories, and also three and  four boson vertices for non--abelian gauge
theories).  For each of these vertices, $Z$ is given as a  product of
different renormalization constants $Z_i$: 
\begin{equation} 
Z =Z_{ex.{}leg}Z_{v}^{-1} Z_3^{\frac{1}{2}} \;\;.
\end{equation}
 Here $Z_3$ is the wave function renormalization constant (WFRC) 
of the relevant gauge boson; 
$Z_{v}$ is the vertex renormalization constant; $Z_{ex.{}leg}$  
are the WFRC of the external legs. The Ward--Takahashi identities guarantee 
the existence of   
relations among the infinite parts of  different $Z_i$, in such a way that the
infinite part of $Z$ does not depend on the vertex we  choose. But, as it is  
well known,  due to the presence  of  masses, this is not true for the finite
part: in  each vertex the masses involved are different. Yet another
problem arises due to  the  gauge dependence of the $Z_i$; although
each $Z_i$ is gauge dependent, their  product, $Z$, cannot depend on the 
choice of 
gauge since \g{r} must be gauge independent. Again, this is true for the
gauge dependence of the infinite part, but it is not  true in  general for the
finite contributions. Therefore, in order to get an acceptable \g{r} in an MDSP
we have  to define $Z$ in a $universal$ (independent of the kind of vertex), 
and gauge--independent way. This is easiest for abelian couplings, 
since in this  
case  one can extend the abelian Ward identity, $Z_{ex.leg}= Z_v$, to  the
finite parts, in such a way that $\g{r}=Z_3 \,\g{0}$, $Z_3$  being 
both gauge independent  and $universal$. This definition of \g{r} is equivalent
to defining an  effective coupling via \cite{qed}: 
\begin{equation}
\frac{1}{\g{eff}}=\frac{1}{\g{0}}+ \Pi_B \;\;,
\end{equation}
where $\Pi_B$ is the transverse {\it bare} vacuum polarization. 

But this  no longer  works in  the broken phase of non--abelian theories. 
Because of  tree level
mixing of  the neutral gauge bosons of $SU(2)$ and $U(1)$, a calculation of
radiative corrections to the proper self energies, $\Pi^{\mu \nu}_{(Z^0)}$ and
$\Pi^{\mu \nu}_{(A)}$,   at 1--loop order, gives a contribution
$\Pi^{\mu\nu}_{(ZA)}$ which is  (i)  not purely transverse and (ii) it further
mixes the self--energies.  In an self--explanatory notation: 
\begin{equation} 
\Pi^{\mu \nu} (q^2)= (g^{\mu \nu} q^2- q^{\mu} q^{\nu}) \Pi^T(q^2) +
g^{\mu \nu} \Pi^L (q^2) \;\;,
\end{equation}
where $\Pi^L(q^2)$ is  in general proportional to  a  mass squared. The 
longitudinal term only  appears when the symmetry is broken, 
since $\Pi^{\mu \nu}$ is purely transverse in an unbroken gauge theory. Then,
if the longitudinal part $\Pi_{ZA}^{L}$  contributes to the mass matrix of
the ($Z^0,{}A$) system,  the  photon would not be massless at 1--loop order!
To maintain a massless photon  without  further field redefinitions one needs  
to eliminate the contribution from  $\Pi_{ZA}^{L}$. 

These  problems affect the couplings \e{} and \g{2}, since they reflect  
the non--abelian nature of the parent gauge theory. Kennedy and  Lynn
\cite{lynn} have shown how to solve them, and,  at the  same time, 
define a \g{r_i} appropriate to phenomenology. They  split  the vertex
contribution for \g{2} into a finite  process--dependent part, and  a
``universal" part $\Gamma^{'}$, consisting of an infinite term  (dictated by
the Ward identities) and a finite term to be  determined\footnote{ From now
on, the term ``vertex"  will  refer to the  contributions due to both the 
vertex itself  and external legs.}.
  As pointed out by these authors, all the  gauge dependence is included in the
universal part. Before breaking the  $SU(2)_L\times U(1)_Y$ symmetry,  the
\g{2} coupling is redefined into  
\begin{equation}
 \tg=\g{2} (1+\g{2} \Gamma^{'})\;\;.
\label{g2tilda}
\end{equation}
When  the symmetry is broken we can write the electric charge in terms of 
\tg and  $g'^2$ as 
\begin{equation}
\frac{1}{\te}=\frac{1}{\tg}+\frac{1}{g'^2} \;\;.
\end{equation}
The advantage of working with the $\tilde{g_{i}}^2$ is now clear, since 
all non--abelian effects are now dumped into $\Gamma^{'}$, 
and one may  define  effective charges associated with the bare tilde 
parameters in exactly the same way as it was done in the  abelian 
case\footnote{ Notice
that the $\Pi^i_B$ refers only to the transverse part}, {\it i.e.} 
\begin{equation}
\frac{1}{\g{i} (q^2)}=\frac{1}{\tilde{g_{i}}^2}+ \Pi_B^{i} (q^2) \;\;.
\end{equation} 
 For the couplings \g{2}, $\e$ we get the effective couplings:
\begin{equation}
\frac{1}{\g{i} (q^2)}=\frac{1}{\g{i}}+ (\Pi_B^{i} (q^2)-2 \Gamma^{'}
(q^2)) \;\;.
\label{effcharge}
\end{equation}
At the same time, if we calculate the 1--loop corrections to the mass matrix 
of the neutral bosons, the non diagonal term will be equal to $\Pi_{ZA}^{L} +
m_W^2  \Gamma^{'}$, and this term is now {\it finite}. The finite part of
$\Gamma^{'}$ may be chosen so that:
\begin{equation}
\Pi_{ZA}^{L} +m_W^2 \Gamma{'} =0 \;\;.
\end{equation}
 
 Thus (a) the photon remains massless (at least to 1--loop) and  (b) we have
available  scheme--independent effective charges\footnote{ It may be shown
that (\ref{effcharge}) remains gauge independent  when coupled to the RGE
for the gauge parameter $\alpha$;  now, since $\alpha=0$ is a fixed point of
this RGE, we will choose to work in the Landau gauge where the value of
$\alpha$  stays at zero, and therefore Goldstone bosons and ghosts remain
massless.}. 

 For the $SU(3)$ coupling the equivalent of (\ref{effcharge}) reads:
\begin{equation}
\frac{1}{\g{3}(q^2)}=\frac{1}{\g{3}}+(\Pi_B^{(3)}(q^2)-2\Gamma^{(3)} (q^2))
\;\;,
\end{equation}
where  $\Pi_B^{(3)}(q^2)$ is the bare proper self--energy of the gluon, 
and $\Gamma^{(3)} (q^2)$ is the universal part of the gluon vertex. Since 
gluons are massless, the process--independent part  of the 
{\it vertex}  consists of an infinite and gauge dependent term, while 
$\Pi_B^{(3)}$ includes the contributions  due to massive quarks.

 The above equations for the effective charges  can be written in terms of
the  couplings at  scale $m_Z$  in the form:
\begin{equation}
\frac{1}{\g{i} (q^2)}=\frac{1}{\g{i} (m_Z^2)}+(\Pi_B^i (q^2)-\Pi_B^i(m_Z^2)
                      -2\Gamma^{'}(q^2)+2\Gamma^{'}(m_Z^2)) \;\;;
\label{effchargez}
\end{equation}
the combinations  $\Pi_B^i(q^2)-\Pi_B^i(m_Z^2)$,
$\Gamma^{'}(q^2)-\Gamma{'}(m_Z^2)$  that appear in (\ref{effchargez}) are now 
finite. Using dimensional regularization we calculate $\Pi_B^i,{}
\Gamma^{'}$; in the Landau gauge we get the following expressions: 
\begin{eqnarray}
(4\pi)^2 \Gamma^{'} &=& \frac{3}{2}\left( \ep - \l{m_W^2} \right)+ 
                         F_{\Gamma} (a_W) \;\;,\\
\label{1}
                    & & \nonumber\\
\Pi^A &=& \Pi^A_{WW}+ \Pi^A_{W+} + \Pi^A_{++}+ \Pi^A_{f} \;\;,\\
(4 \pi)^2 \Pi^A_{WW} &=& -\left\{ \frac{13}{3} \left( \ep- \l{ m_W^2} \right) 
                         +F_1(a_W,a_W)  \right\} \nonumber\;\;, \\
(4 \pi)^2 \Pi^A_{W+} &=& -2 a_W F_2 (a_W,0) \nonumber \;\;,\\
(4 \pi)^2 \Pi^A_{++} &=& \frac{1}{3} \left( \ep - \l{-p^2} + \frac{8}{3} 
                         \right) \nonumber \;\;,\\
(4 \pi)^2 \Pi^A_f &=& \frac{4}{3} \sum_f Q_f^2 \left( \ep- \l{m^2_f} + 
                         F_f(a_f,a_f) \right) \nonumber \;\;,\\
                  & & \nonumber \\
\Pi^W &=& s^2_{\theta} \Pi^W_{WA} +c^2_{\theta} \Pi^W_{WZ} 
         +\Pi^W_{Wh}+ s^2_{\theta}\Pi^W_{A+}
         + s_{\theta}^4 \Pi^W_{Z+}
         +\Pi^W_{h+} +\Pi^W_{2+}+ \Pi^W_f \;\;,\\
(4 \pi)^2 \Pi^W_{WA} &=& -\left\{ \frac{13}{3} \left( \ep-\l{m^2_W} \right)+
          F_1(a_W,0) \right\} \nonumber \;\;,\\
(4 \pi)^2 \Pi^W_{WZ} &=& - \left\{ \frac{13}{3} \left( \ep-\l{m_Wm_Z} \right)+
          F_1(a_W,a_Z) \right\} \nonumber \;\;,\\
(4 \pi)^2 \Pi^W_{Wh} &=& -a_W F_2 (a_W,a_h) \nonumber \;\;,\\
(4 \pi)^2 \Pi^W_{A+} &=& -a_W F_2 (0,0) \nonumber \;\;,\\
(4 \pi)^2 \Pi^W_{Z+} &=& -a_Z F_2 (a_Z,0) \nonumber \;\;,\\
(4 \pi)^2 \Pi^W_{h+} &=& \frac{1}{12} \left( \ep-\l{m^2_h} \right) +
          F_3(a_h,0) \nonumber \;\;,\\
(4 \pi)^2 \Pi^W_{2+} &=& \frac{1}{12} \left( \ep -\l{-p^2} +\frac{8}{3}
         \right)  \nonumber \;\;,\\
(4 \pi)^2 \Pi^W_f &=& \frac{1}{3} \sum_f \left( \ep-\l{m_{f1}m_{f2}}+ 
          F_f(a_{f1},a_{f2}) \right) \nonumber \;\;,\\
                     & &  \nonumber \\
(4 \pi)^2 \Gamma^{(3)} &=& \frac{9}{4} \left( \ep- \l{-p^2} +\frac{4}{3}   
          \right) \;\;,\\
 \Pi^{(3)} &=& \Pi^{(3)}_{gg} +\Pi^{(3)}_f \nonumber \;\;,\\
(4 \pi)^2 \Pi^{(3)}_{gg} &=& - \frac{13}{2} \left(
          \ep-\l{-p^2}+\frac{97}{78}  \right) \nonumber \;\;,\\
(4 \pi)^2 \Pi^{(3)}_f &=& \frac{2}{3} \sum_f \left( \ep- \l{m^2_f} +
          F_f(a_f,a_f) \right)  \;\;,\nonumber 
\label{2}
\end{eqnarray} 
 where we have introduced the following abbreviations  
\begin{eqnarray*}
& &a_i = \frac{m^2_i}{-p^2} \;\;,\;\;s_{\theta}^2 = \sin^2\theta_W 
\;\;,\;\;c_{\theta}^2 = 1-s_{\theta}^2 \;\;,\\
& &\frac{2}{\varepsilon} = \frac{2}{n-4} -\gamma +\ln 4 \pi \;\;.
\end{eqnarray*}
The self--energies above come from the diagrams of Figs.\ (\ref{feyn1}) and
(\ref{feyn2}).
 The sliding  mass scale is denoted by $\mu$, and $n$ is the
dimension of the space--time. Notice that the effective charges do 
$not$ depend on $\mu$, but only on the masses and the external momenta. The 
$F_i$  functions  contain the threshold effects for particle production. The  
functions  $F_i(a_j)$,  $i=1,3,f$, behave like $ A_i\ln a_j$ when $a_j$ goes
to zero, and they tend to a constant value when $a_j$ goes to infinity. On the 
other hand, $F_2(a_j) \rightarrow 0$ in both limits. 
Therefore,  in the high energy limit, $a_i \rightarrow 0$,  we recover
the  $\overline{MS}$--expressions: 
\begin{equation}
\frac{1}{\g{i}(\mu^2)}= \frac{1}{\g{i}(m_Z^2)} + \beta_i \ln 
                        \frac{m_Z^2}{\mu^2} \;\;,
\end{equation}
where, for 3 generations of fermions and 1 scalar doublet
\begin{equation}
(4 \pi)^2 \beta_e = \frac{11}{3}  \;\;,\;\;\;
(4 \pi)^2 \beta_2 = -\frac{19}{6} \;\;,\;\;\;
(4 \pi)^2 \beta_3 = -7 \;\;.
\end{equation}
  Equation (\ref{effchargez}) allows us to study the evolution of the
couplings from $m_Z$ to higher energies, and therefore  to extract
consequences concerning perturbative unification, taking into account all the 
caveats and ambiguities discussed above. 
The input values we will
use, at the scale $m_Z$, are \cite{ellis}:
\begin{eqnarray*} 
\sin^2 \theta_W&=& 0.2329 \pm 0.0013 \;\;,\\
 \al{e} &=& 127.9\pm 0.3\;\;,\\ 
\alpha_{3} &=& 0.111 \pm 0.003 \;\;,\\ 
m_W &=& 80.6\pm 0.4 \,GeV\;\;,\;\;\; m_Z = 91.161 \pm
0.031\,GeV \;\;. \end{eqnarray*}
Due to the  embedding of the SM in $SU(5)$, one  normalizes the $U(1)_Y$
coupling as \cite{quinn}:
$$
\al{1}=\frac{3}{5}{\alpha'}^{-1}=\frac{3}{5} \left( \al{e}- \al{2} \right)\;\;.
$$
  Since the top and the Higgs have  not yet been detected, we will take   
their  masses as free parameters, whose lower experimental bounds will 
be set as :
$m_t \geq 90\;GeV$, $m_h\geq 48\;GeV$ \cite{data}. On the other hand,
these masses can  not be  much higher than about 200 $GeV$  as required by 
perturbative bounds \cite{cabbibo}. This is the
value we will adopt for both $m_t$ and m$_h$;  however, it can be seen that
values between  $100-1000\;GeV$  give rise to the same {\it qualitative}
behavior.
 
 In Fig.\ (\ref{sm1}) we have plotted $\al{i}(\mu)$ calculated with the three
1--loop procedures  previously discussed: mass dependent (MD) procedure (eq.
(\ref{effchargez})),  integration with $\theta$ functions, and $\overline{MS}$. 
Only  the high energy region $( 10^{12}-10^{19}\;GeV)$, where
unification\footnote{Of course this is not the case for  the SM. This plot is
simply intended to illustrate how the   decoupling affects the evolution of
the coupling constants.}  
could occur, is depicted.  We see at once that the
approximation we use for \al{1} and \al{3} make  some difference although
in this case very little: for example, \al{3}--MD is  a little larger than the
other couplings, due to the fact that when thresholds are taken  into
consideration the top quark decouples at low energies. This effect
propagates to high energies by the RGE and  in this case the decoupling is 
smoother than in the $\theta$ approximation. 

 In the running of \al{2} a larger number of massive particles participate, 
and therefore their decoupling will introduce larger corrections than for the 
other two couplings. This is readily seen in the figure, which also shows that 
\al{2}--$\theta$ is a little larger than \al{2}--$\overline{MS}$ (same reason
as for \al{3}), and they both are larger than  \al{2}--MD.
We now  have the contributions of $W^{\pm}$ and $Z^0$, which in the
mass dependent scheme take  ``longer" to decouple. This is  the dominant
decoupling effect in \al{2},  even though $m_Z,{} m_W\leq m_t,{} m_h$, which 
is not operative  in \linebreak 
\al{2}--$\theta$ because we begin to integrate
precisely at the scale of $m_Z$. That is, the thresholds effects tend to make 
\al{2} ``less asymptotically free". This was already pointed out in Refs.
\cite{ross1}, \cite{ross2} which show that, within the context of the 
minimal $SU(5)$ model, the $\beta$--function  for the  $SU(2)$ coupling
with thresholds is  positive in the region around $m_W$. 

We have also calculated \al{i} at 2--loop order $without$  thresholds. The
2--loop RGE's in the $\overline{MS}$  are well known \cite{jones}; 
solving these equations  by an iterative technique \cite{hall,marciano}, 
one  obtains:
\begin{eqnarray}
\label{2loop}
 \al{i}(\mu)&=& (\al{i}(\mu))^{(1)}+\frac{b_{ij}}{(4\pi) b_j} \ln
             \frac{\alpha_j(m_Z)}{\alpha_j(\mu)} \;\;,\\
\al{i}(\mu)^{(1)} &=& \al{i}(m_Z)+ \frac{b_i}{(4 \pi)} \ln 
             \frac{m^2_Z}{\mu^2} \nonumber\;\;,
\end{eqnarray}
where $i,\,j=1,\,2,\,3$; the coefficients $b_i,\;b_{ij}$ for 3  generations
of  fermions and 1 scalar doublet are: 
\begin{equation}
b_i= \left( \begin{array}{c} 41/10\\ -19/6 \\ -7 \end{array}
\right) \;\;,\;\;\; 
b_{ij}= \left( \begin{array}{ccc} 199/50 & 27/10 & 44/5 \\ 
                                  9/10   & 35/6  & 12   \\
                                  11/10  & 9/2   & -26  
               \end{array} \right) \;\;.
\end{equation}
In this case we calculate directly \al{1}, rather than \al{e}, because
working in $\overline{MS}$ is equivalent to working with the symmetric theory 
$SU(3)\times SU(2)\times U(1)$.

In Fig.\ (\ref{sm12}) we can see that 2--loop effects are qualitatively
similar to  MD effects:  
they raise \al{3} and decrease \al{2},  even though in  the
case of \al{2}  the  effect from the decoupling of the gauge bosons
dominates over 2-loop effects. Thus, if we take into account both of them 
(2--loop and thresholds) by calculating \al{i} with an MDSP at 2--loop
order, \al{2} will be less asymptotically free that \al{2}--$\overline{MS}$ 
at 1--loop.

Short of doing the exact calculation we have studied this case with 
an appropriate 
approximation for the threshold functions. We approximate 
$\al{i}^{(1)}$, $b_i,\;b_{ij}$  in Eq. (\ref{2loop}) by: 
\begin{eqnarray}
\al{i}(\mu)^{(1)} &=& \al{i}(m_Z) + \frac{1}{(4\pi)} 
\sum_k b_i^{(k)}
\left( f^{(k)}(a_k) \mid_{(-p^2=m_Z^2)}-f^{(k)}(a_k) 
\mid_{(-p^2=\mu^2)} \right) \nonumber \;\;,\\
                       & & \\
b_i &=& \sum_k b_i^{(k)} f'^{(k)}(a_k) \;\;, \\
b_{ij}&=& \sum_l\sum_k b_{ij}^{(k,l)} f'^{(k)}(a_k) f'^{(l)}(a_l)
\;\;,
\label{bij}
\end{eqnarray}
where we have summed over all the particles in the SM, and \cite{georgi}: 
\begin{eqnarray}
f^{(k)}(a_k)&=&\ln (1+c_k a_k) - \ln c_k a_k \;\;,\\
f'^{(k)}(a_k) &=& \frac{d\,f^{(k)}}{d \ln (-p^2)}= 
\frac{1}{1+c_k a_k}\;\;. 
\end{eqnarray}
These functions $f^{(k)}$ and $f'^{(k)}$ have been chosen so that their
behavior in the limit $a_k\rightarrow 0$ is:
\begin{equation}
f^{(k)}(a_k) \rightarrow-\ln c_k a_k\;\;,\;\;\; f'^{(k)}(a_k) \rightarrow 1
\;\;, 
\end{equation}
and both functions vanish when $a_k \rightarrow \infty$. Thus in the high
energy limit (or  for massless particles) we  recover the 
\al{i}--$\overline{MS}$ 
expressions; the heavy masses decouple at low energy more smoothly than in 
the decoupling model by a step  function. This approximation is very good
at 1-loop order: comparing the  numerical results with those obtained from
\al{i}--MD,  the differences are only of order  0.1\%. We also 
expect this to be true at 2--loop order. Actually it is not necessary to  
modify the 2--loop coefficient $b_{ij}$, since there are no appreciable
differences if we take  $b_{ij}$ as given by the $\overline{MS}$  prescription
or as given by  Eq.(\ref{bij}). This may suggest that  2--loop thresholds are
not relevant  (only 1--loop thresholds are). 
In fact, 2--loop threshold effects are quite small since they are
corrections to the 2--loop $\overline{MS}$ contribution which is, in turn, a 
correction to the 1--loop contribution.

If we compare the 1--loop and 2--loop results, both within this
approximation,  we see  (Fig.\ \ref{sm12a}) that ``2--loop+thresholds" has the
effect of lowering \al{2} and  raising \al{3} . Therefore, combined effects of
mass and order of the perturbation theory, tend  to bring \al{2} and  \al{3}
$closer$ at high energies.  Unfortunately this effect is  not strong enough to
unify the three couplings of the SM, but it is interesting to point out its
existence.

 It is clear this effect will become the more relevant the larger the number
of  massive particles we have.  So far we have only worked with the
standard  matter content of the SM, in which  all the particles have been
detected, and their masses measured, except for the  top and the Higgs. The
particle contents can be modified,  for example by extending the SM and
working  in the so called Minimal Supersymmetric Standard Model (MSSM).
Here more degrees of freedom (the susy partners etc.)  begin to
contribute between $m_Z$ and the putative unification scale.  Since  these
susy particles have not been detected,  there are at most lower
bounds available to their masses. We can make use of  some relations between
them which  reduce the number of the necessary arbitrary mass parameters
\cite{susy}.  
The common mass for both squarks and sleptons, $m_0$, and the gaugino mass, 
$m_{1/2}$, may be bounded by demanding that
susy masses be in the range  between $45\; GeV$ and $1\; TeV$. 
We also take  $m_Z \leq  m_{\mu},\;m_+\simeq m_H \leq 
1\;\;  TeV $, and  $m_t,\;m_h \simeq 200 \;  GeV $, where $m_{\mu}$ is the 
higgsino mass.

 Previous works in MSSM indicate that an $M_X\approx  
O(10^{16})$  and a common $M_{Susy}$ (or explicit susy masses) 
in the range from $m_Z$ to $O( 1 TeV)$ are compatible with experimental bounds 
on proton decay \cite{ellis,amaldi,anselmo}. In these analyses, and 
except for the case of 1--loop $\overline{MS}$, the  values  for  
$\alpha_3(m_Z)$ larger than 0.111 are favoured to have an $M_{Susy}$ 
in this range. 
The latest data on $\alpha_3(m_Z)$ from LEP (see table 1) \cite{data3} 
indicate an $\alpha_3(m_Z)$ larger than this by 12\%. Thus, until we  have a 
more precise determination of $\alpha_3(m_Z)$ we will have to take the 
allowed  range
for $\alpha_3(m_Z)$ to be (0.108, 0.125).  Systematically one sees that the 
higher the value of $\alpha_3(m_Z)$ the  higher $M_X$,  but  the  lower
$M_{Susy}$.  This systematic also happens when we use  2--loop instead of 
1--loop RGE's \cite{anselmo}. The trend is maintained  using Mass 
Dependent RGE's,  but $now$ the values of the susy masses  needed to unify
the couplings  are $higher$ than with the other methods.

In Fig.\ (\ref{susy}) we have represented \al{i}--MD for two different values
of the susy  parameters and $\alpha_3(m_Z)= 0.111$. We have also included
\al{i}--$\overline{MS}$ with  $M_{Susy}=m_Z$ and the same value of 
$\alpha_3(m_Z)$ in order to compare the results yielded by different  
approximations. As pointed out before, the \al{i}--$\overline{MS}$  
unify, while the 
\al{i}--MD do not. For example, we would need  susy masses of order  
$10^2\; TeV$ for $\alpha_3(m_Z)=0.111$, and  of order 1 $TeV$ for 
$\alpha_3(m_Z)=0.120$. 
With $\alpha_3(m_Z)$ in the range given before, and requiring 
$M_X\simeq O(10^{16})$ to avoid conflicts with proton decay, susy masses of
order $m_Z$ are excluded, except  $m_{1/2}\simeq m_{\tilde{w}}$ which has
to be less than $O(3\, TeV)$. At 2--loop order, with  complete light
thresholds at 1--loop, the bounds increase:  $m_{1/2}$ greater
than approximately $3.5 \,TeV$  is excluded,  but the rest of  susy masses
must be greater than $O(10\, TeV)$.

So far, we have discussed perturbative unification within MSSM, but without
any reference to the unification group $G$, and the new heavy fields which
are  introduced by $G$ in the theory. Moreover, we have required that the
couplings  ``unify at the scale $M_X$" when  they  really ``cut at the scale
$M_X$"  and run separately after it. To speak properly about perturbative
unification, we need that  the couplings converge to only a single running
coupling constant: the $G$ coupling.  We would obtain this when we work
within $G$, the group $SU(5)$,  taking into account the thresholds of all the
masses, both light and heavy masses \cite{ross1,anselmo2}.  As we have
seen, light thresholds introduce appreciable differences in the  running of the
couplings, and the same effect takes place with the heavy  masses when
the scale approaches  $M_X$ \cite{propia}. Although we will have many 
more free parameters (the heavy masses), the trend is the same that in the 
simplest case when no reference was made to $G$: with mass dependent
RGE's we need  heavier susy masses in order to obtain unification with
$\alpha_3(m_Z)$  in the range allowed by experimental data. 

In conclusion, we have studied the evolution of the running couplings of the 
SM including complete threshold effects due to the light particles. 
We have also used effective charges  \cite{lynn}, which are {\it mass
dependent}, and {\it process} and {\it gauge} independent when we choose an
adequate initial  condition for the gauge parameter (Landau gauge). We have
also seen that the effect  of the decoupling of the light masses is $not$ 
negligible at high energy. We apply the same method to study the MSSM, and 
the unification of the couplings within this model. Finally, We have found 
that light thresholds have the property of raising the values of the susy 
masses needed to keep  $\alpha_3(m_Z)$ within the  range of the already
available experimental values.

\noindent{\bf Acknowledgments}

One of the authors, M. B-G., would like to thank Dr. L. Garay for a critical 
reading of the manuscript.
\pagebreak
\pagebreak
\begin{table}
  \begin{tabular}{|c|c|c|}
\hline
 Experiment & Central value & Error  \\
\hline
    ALEPH jets & 0.125         & $\pm 0.005$    \\
    DELPHI jets& 0.113         & $\pm 0.007$    \\
    L3 jets    & 0.125         & $\pm 0.009$    \\
    OPAL jets  & 0.122         & $\pm 0.006$    \\
    OPAL $\tau$& 0.123         & $\pm 0.007$   \\
\hline 
   $J/\Psi$    & 0.108         & $\pm 0.005$   \\
   $\Upsilon$  & 0.109         & $\pm 0.005$    \\
   Deep Inelastic& 0.109       & $\pm 0.005$    \\   
\hline
  \end{tabular}
\caption[]{Experimental values of $\alpha_3(m_Z)$}
\end{table}
\pagebreak
\figure{ The 1--loop particle contributions to the $\Pi^{(j)}$
functions (j= $A$; $W^{\pm}$). (a) Pure gauge boson contributions plus ghost
($\omega_i$) contributions. (b) Graphs that mix scalars and gauge bosons.
(c) Pure scalar contributions. (d) Fermion contributions. $\phi_1$ 
stands for the physical Higgs scalar; $\phi^{\pm}$, $\phi_2$ for the
Goldstone bosons associated with $W^{\pm}$ and  $Z$; $f$ for fermions (left
and right); and $f_{1L}$, $f_{2L}$ for the components of each doublet of left
fermions. \label{feyn1}}
\figure{ The 1--loop particle contributions to the $\Pi^L_{ZA}$ 
function, $i.e.$ to $\Gamma^{'}$. (a) Gauge boson plus ghost. (b) Scalars and
gauge bosons. \label{feyn2}} 
\figure{ Evolution of the three couplings of the SM calculated
with the three 1--loop procedures: mass dependent (solid lines), $\theta$
function (dashed lines), and $\overline{MS}$ (dotted lines). \label{sm1}}
\figure{ Evolution of the three couplings of the SM at 2--loop
order with $\overline{MS}$ (solid lines), and 1--loop order with a mass 
dependent method  (dashed lines). 
We also plot 1--loop $\overline{MS}$ for comparison (dotted
lines).\label{sm12}} 
\figure{ Evolution of the three couplings of the SM at 1--loop
order (solid lines), and 2--loop order (dashed lines), both with approximated
threshold functions.\label{sm12a}} 
\figure{ Evolution of the three couplings of the MSSM at 1--loop
order calculated with (a) $\overline{MS}$ with $M_{susy}=m_Z$ (solid lines); 
(b) MD with $m_{1/2}= 45\; GeV$ and  $m_0=m_Z$ (dashed lines); 
(c) MD with $m_{1/2}=m_0=1\;TeV$ (dotted lines); 
we take $m_{\mu}=m_{+}=m_H=m_0$ and $m_t=m_h=200 \; GeV$. \label{susy}}

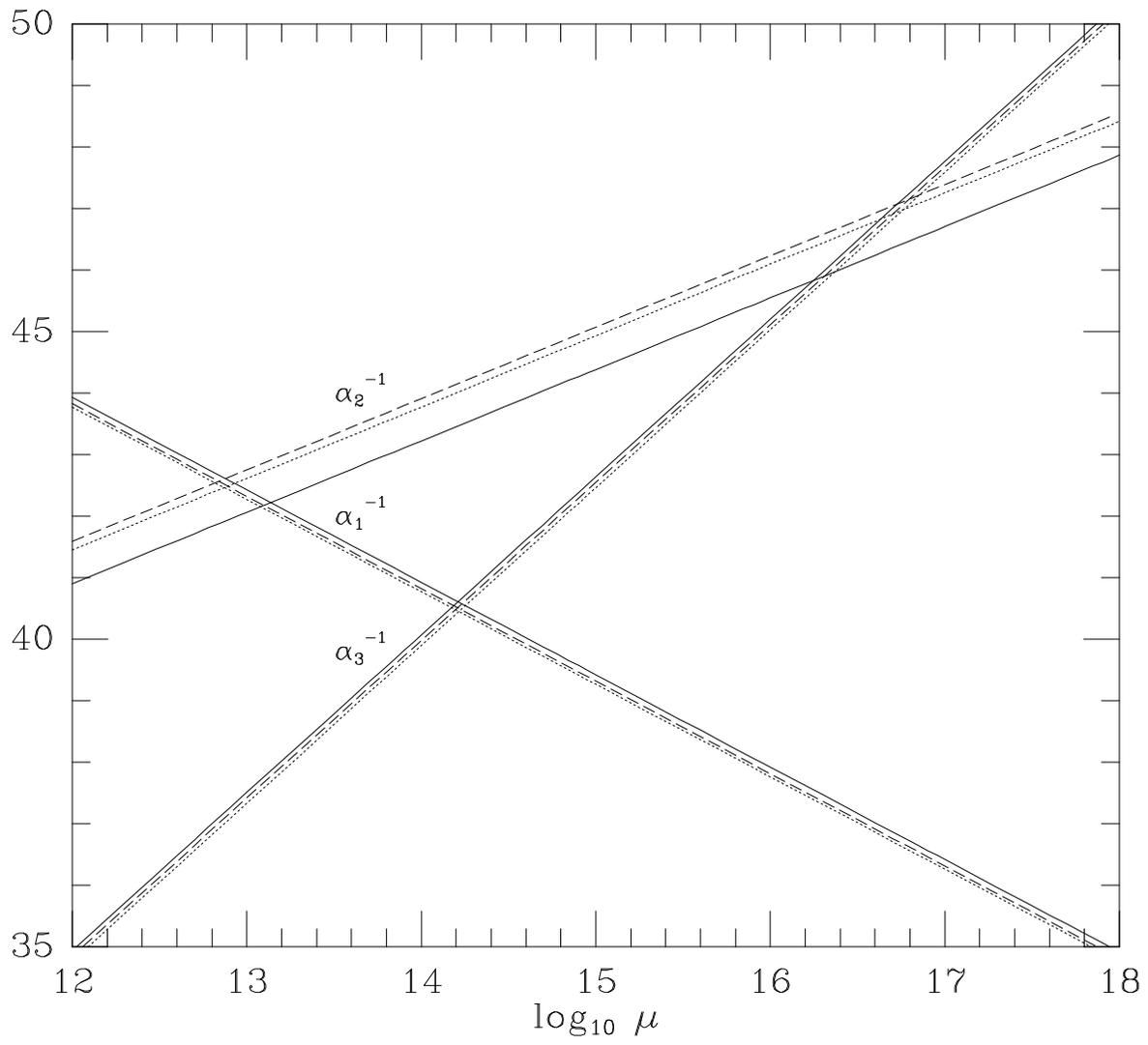
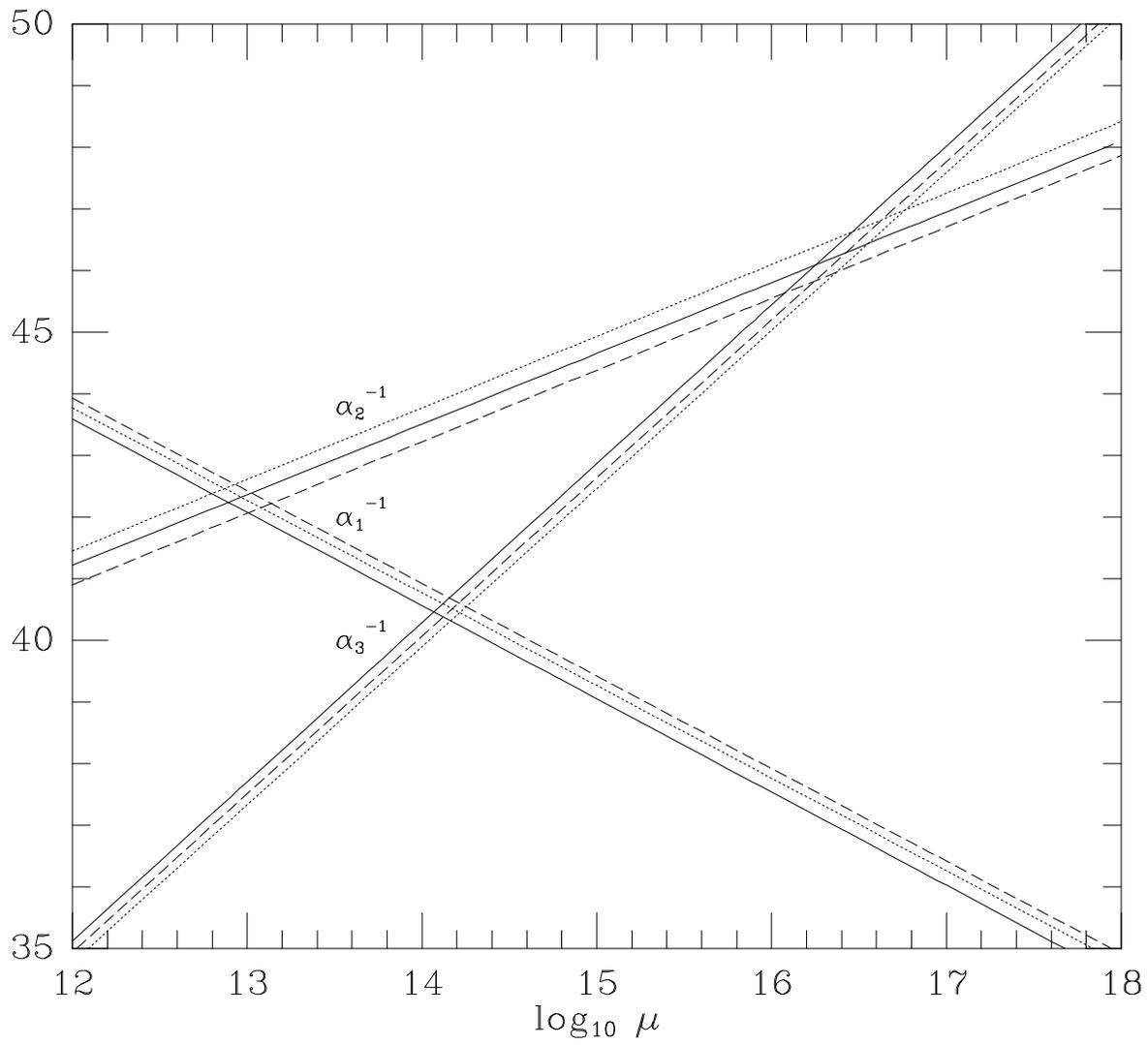
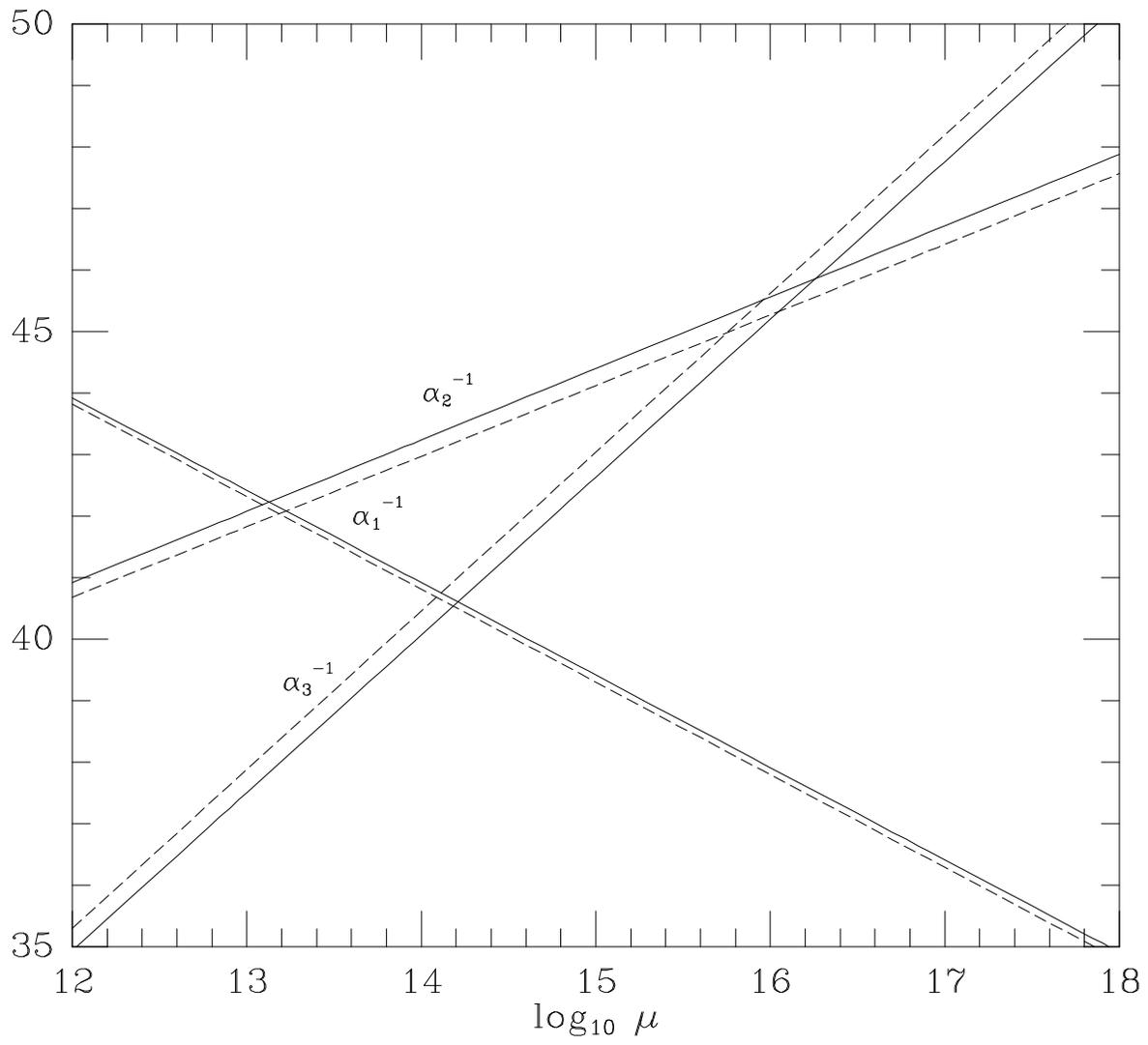
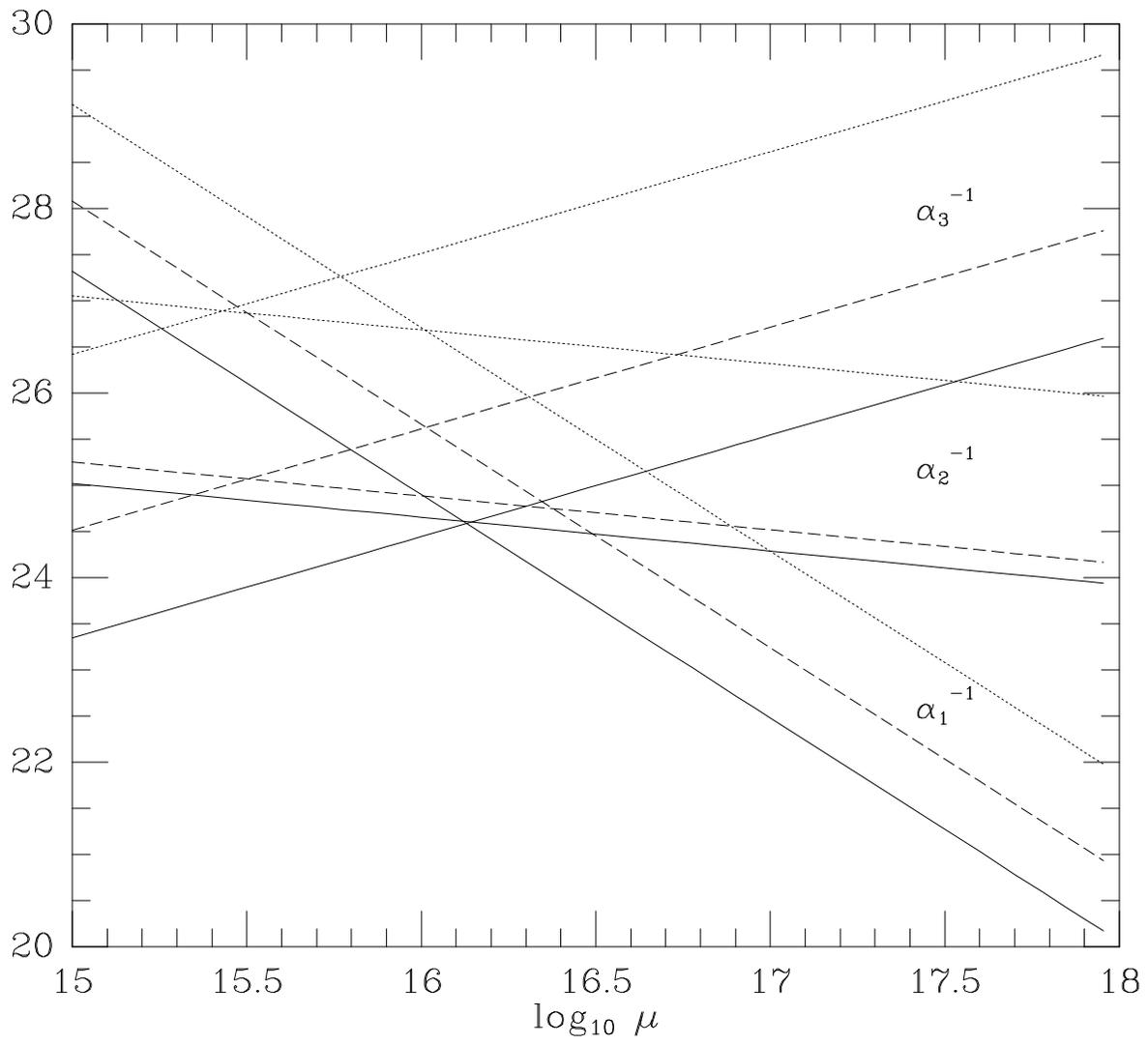
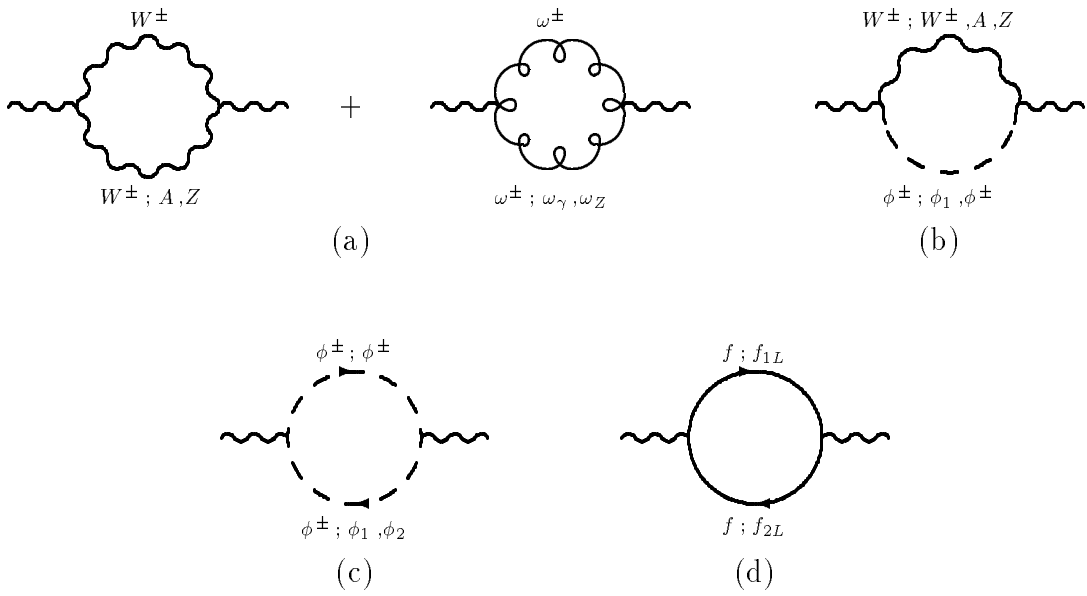
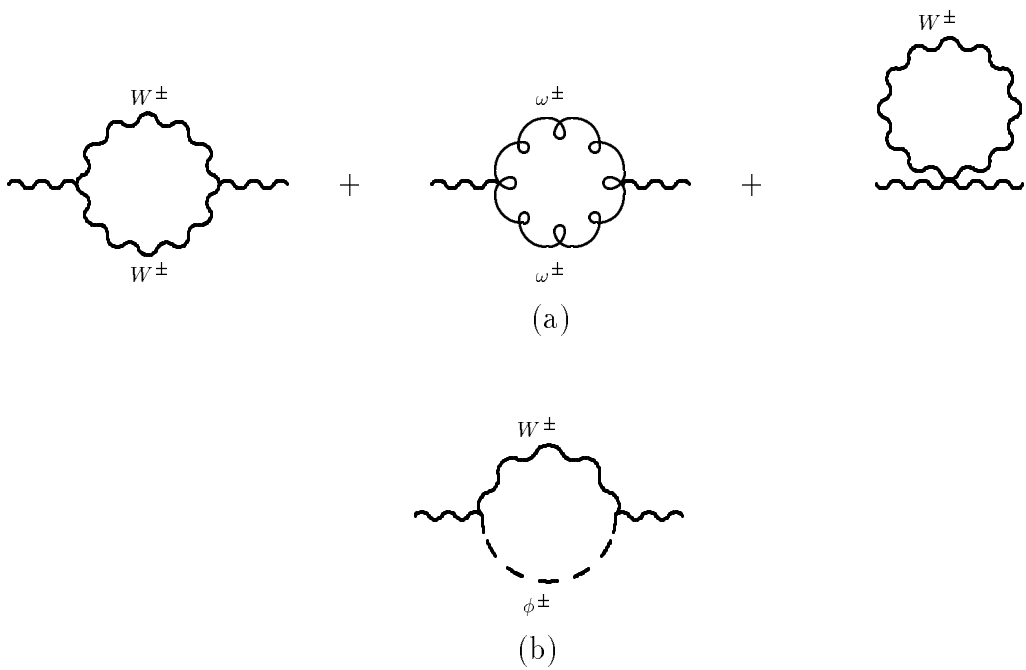
\end{document}